# Polarization-sensitive and broadband photodetection based on a mixed-dimensionality TiS₃/Si p-n junction

Yue Niu[1,2], Riccardo Frisenda[2]*, Eduardo Flores[3], Jose R. Ares[3], Weicheng Jiao[1], David Perez de Lara[2], Carlos Sanchez[3,4], Rongguo Wang[1], Isabel J. Ferrer[3,4]* and Andres Castellanos-Gomez[5]*

[1]Science and Technology on Advanced Composites in Special Environments Laboratory, Harbin Institute of Technology, 150080, Harbin, P. R. China

[2]Madrid Institute for Advanced Studies (IMDEA Nanociencia). Ciudad Universitaria de Cantoblanco Calle Faraday 9, 28049 Madrid, Spain
E-mail: riccardo.frisenda@imdea.org

[3]Materials of Interest in Renewable Energies Group (MIRE Group). Dpto. de Física de Materiales, Universidad Autónoma de Madrid, UAM, Campus de Cantoblanco, E-28049 Madrid, Spain.
E-mail: isabel.j.ferrer@uam.es

[4]Instituto Nicolás Cabrera, Universidad Autónoma de Madrid, UAM, Campus de Cantoblanco E-28049 Madrid, Spain.

[5]Materials Science Factory, Instituto de Ciencia de Materiales de Madrid (ICMM-CSIC), Madrid, E-28049, Spain.
E-mail: andres.castellanos@csic.es

Abstract:

The capability to detect the polarization state of light is crucial in many day-life applications and scientific disciplines. Novel anisotropic two-dimensional materials such as TiS₃ combine polarization sensitivity, given by the in-plane optical anisotropy, with excellent electrical properties. Here we demonstrate the fabrication of a monolithic polarization sensitive broadband photodetector based on a mixed-dimensionality TiS₃/Si p-n junction. The fabricated devices show broadband responsivity up to 1050 nm, a strong sensitivity to linearly polarized illumination with difference between the two orthogonal polarization states up to 350 % and a good detectivity and fast response time. The discussed devices can be used as building blocks to fabricate more complex polarization sensitive systems such as polarimeters.

## 1. Introduction

Polarization sensitive photodetectors, which are optoelectronic devices sensitive to the polarization of the incoming light, have a strong impact in many different branches of science and technology like astronomy,[1] quality assessment in food and mechanical industry,[2] ellipsometry.[3] Natural evolution endowed certain animals and insects with eyes equipped with monolithic polarization sensitive photodetectors. These specialized light detectors facilitate the identification of their preys or allow them to navigate recognizing the polarization patterns of the sky.[4] Interestingly, similar navigation techniques based on the detection of polarization patterns of the sky have been probably used by ancient civilizations during human history.[5] The recent isolation of strongly dichroic 2D semiconductors holds a great promise for the fabrication of monolithic polarization sensitive photodetectors and polarimeters.[6] Up to now, however, most of the reported works on photodetectors based on 2D semiconductors are focused on transition metal dichalcogenides or other chalcogenides with marked in-plane isotropic properties and thus the built-up photodetectors are insensitive to polarized light.[3a, 7] The amount of works exploring the use of dichroic 2D semiconductors is still scarce and they are mostly limited to devices with low detectivity values or with a limited cut-off wavelength.[8] Recently, Island *et al.* have fabricated photodetectors using $TiS_3$,[7c] a member of the trichalcogenide family that presents remarkable quasi-1D electrical and optical properties with an anisotropy larger than the one of black phosphorus.[9] However, this strong in-plane anisotropy has not been exploited yet to fabricate polarization sensitive photodetectors.

Here we demonstrate the fabrication of monolithic polarization sensitive photodetectors based on $TiS_3$. The devices are built up by staking a $TiS_3$ ribbon onto p-type silicon, thus forming a p-n junction based photodiode geometry which allows one to operate the device either in photovoltaic (PV) mode, without applying any external bias, or in photoconductive (PC) mode, with a positive or negative bias applied to the device. In PV mode upon illumination, the built-in electric field present at the interface between $TiS_3$ and Si separate the photogenerated charge carriers yielding to a photocurrent even under unbiased condition. The fabricated devices show broadband photoresponse from 405 nm to 1050 nm and a strong dependence of their photoresponse on the polarization direction of the illumination. The photocurrent generated for light polarized along the *b* axis direction of the $TiS_3$ lattice is up to 350% larger than that polarized along the *a* axis direction. These devices can be used as building blocks to fabricate more complex thin polarization sensitive systems such as polarimeters.

## 2. Results and Discussion

**Figure 1**a-b show optical images of synthesized $TiS_3$ microcrystals with a characteristic ribbon-like shape.[10] **Figure 1**c shows a representation of the crystal structure of $TiS_3$, a monoclinic $ZrSe_3$-type lattice with two lattice vectors per unit cell. The atoms are separated in layers that are held together by van der Waals interactions and each layer is composed of parallel covalently bonded chains. Each Ti atom is bonded to six S atoms belonging to the same chain and two additional S atoms in the neighbor chains.[9a, 9b, 11] The parallel chains are responsible for the quasi-1D nature of the material which leads to an anisotropy in the electrical conductivity and optical properties between the in-plane *a* and *b*-axes.

**Figure 1**d shows various micro-transmittance spectra acquired on a $TiS_3$ ribbon while varying the relative orientation between the linearly polarized incident light and the *b*-axis of the $TiS_3$ ribbon.[12] One can clearly see the gradual evolution of the transmittance spectra using light with polarization angles ranging from aligned along the *a* direction to the $b$ direction. The transmittance along the $b$-axis is quite low (52% at 2.32 eV) and the ribbon appears darker due to the larger absorption. In contrast the transmittance along the $a$-axis reaches 71% at 2.32 eV and the ribbon becomes much more transparent. (See a movie of the polarization dependent transmittance in the Supplementary Information). **Figure 1**e shows a polar plot of the transmittance at different wavelengths as a function of the orientation between the linearly polarized light and the $b$-axis. The minimum (maximum) of transmittance occurs for light polarized in the direction the $b$-axis ($a$-axis). This 2-fold polarization angle dependent absorption feature can be explained by the strong difference in the electrical conductance about the $b$-axis (highly conductive) and the $a$-axis (poorly conductive) observed in TiS$_3$, analogously to what happens in wire grid polarizers.[9] In fact, recent *ab-initio* calculations based on the Bethe-Salpeter equation show that the transmittance and the absorption coefficient of TiS$_3$ show a marked 2-fold polarization angle dependence.[9b,9c] Moreover, in order to proof that this optical spectra anisotropy steems from the anisotropic crystal structure of the TiS$_3$ and not from a geometrical effect (e.g. nanowire wave guiding) we show in the Supporting Information a similar transmisstion spectra measurement on a TiS$_3$ sample with very low aspect ratio displaying very similar features.

To fabricate the TiS$_3$ based photodetectors we employ mechanical exfoliation combined with deterministic transfer of ultrathin TiS$_3$ ribbons.[13] **Figure 2** summarizes the steps carried out to fabricate the devices. The commercial substrate is based on pre-patterned Cr/Au electrodes evaporated on SiO$_2$/Si (Ossilla®). We mask the electrodes using a piece of kapton tape and we etch away the unmasked SiO$_2$ using a glass etchant paste (ArmourEtch®) that effectively removes the SiO$_2$ without damaging the Si bottom layer. Note that this is a very convenient way to etch the SiO$_2$ without employing extremely dangerous buffered HF etching recipes. Finally, the device is completed by transferring a TiS$_3$ ribbon by means of an all dry deterministic transfer method bridging one of the Au/Cr electrodes to the exposed Si surface.[14]

The optoelectronic performances of the fabricated TiS$_3$ devices are characterized in dark and upon illumination using a homemade scanning photocurrent system.[15] **Figure 3**a shows the *IV* characteristics of a TiS$_3$ device in dark and upon illumination (with 660 nm of wavelength) with increasingly optical power. In the dark state the device shows a marked rectifying *IV* due to the p-n junction formed at the interface between the TiS$_3$ (n-type semiconductor) and the silicon (p-type, boron doped with a resistivity of 0.0005 - 0.001 Ohm·cm). The rectifying *IV* can be modeled by the Shockley diode equation (see Supporting Information) as shown in the inset of **Figure 3**a. The fit reproduces well the experimental *IV* when considering a series resistance of 270 kΩ (the fit parameters are listed in Table S1 of the Supporting Information). This series resistance takes into account the contributions of the TiS$_3$ internal resistance and the contact resistance between TiS$_3$ and Au.

Under external illumination we observe an increase in the current. The reverse current (current at negative bias) increases monotonically upon increasing the incident power. **Figure 3**b shows the details of the *IVs* at low bias voltage. At zero bias voltage, the built-in electric field at the interface between the $TiS_3$ and the silicon separates the photogenerated electron-hole pairs giving rise to the so called short circuit current ($I_{sc}$), a clear signature of the PV photocurrent generation mechanism. The open circuit voltage ($V_{oc}$) is also a signature of the built-in electric field. Both $I_{sc}$ and $V_{oc}$ increase as the incident optical power increases (see the Supporting Information for more details).

**Figure 3**c shows the photo-responsivity of the device measured at -0.5 V by employing illumination sources with different wavelengths. The photo-responsivity, $R$, is a measure of the electrical response to light and is defined as

$$R = I_{ph}/P_{in} \qquad (1)$$

where $P_{in}$ is the incident optical power. The shape of the responsivity spectrum follows closely that of the unpolarized micro-absorbance spectrum of the $TiS_3$ ribbon, calculated from the transmittance according to the formula

$$A = -\log{(T)} \qquad (2)$$

suggesting that the photoresponse is dominated by the interaction of light with the $TiS_3$ itself. These results are in good agreement with the previously reported experimental values obtained from photocurrent response and optical absorbance measurements of $TiS_3$.[16] Besides, to evaluate the performance of $TiS_3$/Si as a photodetector, we calculate the inferred detectivity ($D^*$), which represents the ability to detect weak optical signals. Assuming the noise to be limited by shot noise, $D^*$ is calculated from the measured responsivity and the dark current[17] according to

$$D^* = R \cdot A_d{}^{1/2} /(2eI_d)^{1/2} \qquad (3)$$

where $A_d$ is effective area of the device, $e$ is the electron charge and $I_d$ is the dark current. The calculated $D^*$ value reaches $2.5 \cdot 10^8$ Jones at bias voltage of 0 V and $2.65 \cdot 10^9$ Jones at - 0.5V for the device shown in the main text (other devices reach up to $2.65 \cdot 10^9$ Jones and $1.3 \cdot 10^{10}$, respectively). The summarized photonic parameters ($R$, $D^*$, response time) and comparison with reported anisotropic photodetectors are shown in Table 1. Consequently, the superior efficiency of the detectivity of our $TiS_3$/Si devices is comparable to black phosphorus/$WSe_2$ heterojunction[17b] and one or two magnitude larger than other reported anisotropic photodetectors,[8b, 8c] which we attribute to the strong intrinsic responsivity of $TiS_3$ together with the photodiode device architecture that minimizes the dark current.[7c] However, Tan *et al.* reported a photodetector based on GeS shows a rather high detectivity but operating at high temperature.[17a] Also, the devices like GeS and $ReS_2$ have slower response times and limited cut-off wavelength of 800 nm.[8e, 17a]

The electrical and photovoltaic properties manifested by the $TiS_3$-Si device can be explained by the band diagram of the two materials schematically depicted in **Figure 3**d.

The isolated materials (left panel) are semiconductors with bandgap energies of 1.10 eV and 1.12 eV respectively for $TiS_3$ and Si. The difference in the position of the Fermi levels of the isolated materials induces a charge transfer that creates a built-in electric field, when the two materials are brought in contact. This phenomenon is shown in the band diagram by the bending of both the conduction and the valence band at the interface between the two materials (see right panel of **Figure 3**d). The rectifying *IV*s and the photovoltaic effect observed in our devices are consistent with such a band diagram.

Taking advantage of the high-speed optical communication and high responsivity of our devices, we demonstrated a single-pixel camera[18] based on an ultrathin $TiS_3$/Si photodetector. A scheme of the single-pixel camera setup integrated with our $TiS_3$/Si devices is shown in **Figure 4**a.[12] Briefly, the target to image is mounted on a motorized XY stage, which can be controlled by the computer. The target is illuminated from the bottom by a white light source. An optical fiber, positioned on top of the target image, acts as a pinhole, thus only collecting the light coming from a small region of the target. The free end of the fiber is connected to a cannula to irradiate directly the $TiS_3$/Si photodetector reversely biased. Whilst illuminating the sample, the device electrical properties are measured with a Keithley 2450 source meter unit as a function of the spot position. From the spatial map of the photocurrent we can reconstruct a photograph of the target where optically dark areas correspond to low photocurrent signals and bright areas to large photocurrent. The data acquisition and motion control are managed through a home-made routine written in Matlab®. To test the effectiveness of the camera, a "Smiley Face" was used as a target and the recovered image with a resolution of 64 · 64 pixels is shown in **Figure 4**b. In spite of the simplicity of the single-pixel camera setup, it provides a testbed to explore the performance of photodectectors based on nanomaterials in real-life applications.

After testing the optoelectronic properties of $TiS_3$/Si junctions for unpolarized light, we now discuss the performances of these devices for polarized light detection. **Figure 5**a and 5b show the *IV* curves of the device upon illumination with linearly polarized light at 660 nm of wavelength and optical power of 0.54 µW. The polarization of the illumination has been varied from linearly polarized along the *b*-axis (labelled as 0° in the figure) to linearly polarized along the *a*-axis (labelled as 90° in the figure). From the *IV*s, it is clear that the photoresponse of the device depends on the relative orientation between the linearly polarized illumination and the crystal orientation of the $TiS_3$ ribbon, as expected from the strong linear dichroism observed in the absorbance of the material. The inset shows the relationship between the photodetector responsivity (extracted at 660 nm) and the relative angle between the incident linearly polarized light with respect to the *b*-axis. **Figure 5**c and 5d show two polar plots with the polarization orientation dependence of the photocurrent extracted at -2 V (mainly due to photoconductive photocurrent generation) and short circuit current at 0 V (due to photovoltaic effect). The anisotropy in the photocurrent response stems from the anisotropic crystal structure of $TiS_3$. In fact, as shown above, the absorption of photons polarized along the *b*-axis is stronger than that for photons polarized along the *a*-axis. The maximum photocurrent, corresponding to a photo-responsivity *R* of ~ 35 mA / W, is reached for incident light polarized along *b*-axis. The minimum response is ~ 19 mA / W with incident light polarized along *a*-axis. As shown in Table 1, this polarization photosensitivity values are comparable to those of recently reported black phosphorus based polarized photodetectors.[8b, 17b] These figure of merits found for the $Ti_3/Si$ photodetector device are comparable with those of commercially avalaible devices. Specifically, *R* is very close to the responsivity of commercial Si-based photodiodes (Thorlabs S120C, *R* = 55mA/W). Moreover, the sensitivity of $TiS_3$ to the polarization of the incident light allows to fabricate more compact polarization sensitive photodetectors compared to conventional Si photodetectors where there is the need of using an external polarizer element. The maximum anisotropy ratio of the photocurrent (measured at -0.5 V) and short circuit current,

$$\bar{\sigma} = (I_{max} - I_{min}) / (I_{max} + I_{min}) \qquad (4)$$

of the device is 0.29 and 0.64, respectively, which is comparable with the angular dependent photocurrent in other phototransistors (from 0.3 to 0.96).[8, 17, 19] However, those polarization sensitive photodetectors typically require an external bias as the driving force to prevent the recombination of photo-generated electron-hole pairs and their dark current is larger (thus typically yielding to lower detectivity values). We address the reader to the Supporting Information to see the full characterization of another three $TiS_3/Si$ photodetector devices.

**Figure 6** shows a spatially resolved characterization of the device photoresponse. The incident light at 660 nm with the power of 0.54 µW was applied (the motor step size is 3 µm). **Figure 6**a shows an optical image of the studied device region. **Figure 6**b shows four photocurrent maps acquired rotating the relative orientation between the linearly polarized light and the $TiS_3$ b-axis. The different regions of the device, outlined in the photocurrent maps, are determined from reflectivity maps acquired simultaneously (see Supporting Information). Interestingly, from the photocurrent maps one can clearly observe how the photocurrent is only generated at the $TiS_3/Si$ junction, which is because of the charge separation happens right at the heterojunction region.

## 3. Conclusion

In summary, we presented a self-powered polarization sensitive photodetector based on a $TiS_3/Si$ p-n junction. The heterostructure device demonstrates a broadband photodetecting range (from 405 nm to 1050 nm) with high photo-responsivity (~ 35 mA/W), but also shows a high sensitivity to polarized infrared illumination photodetection. The polarized contrast between *b* and *a*-axis direction of the $TiS_3$ lattice is up to 350%. Our novel heterojunction structure offers an ideal candidate for future 2D optoelectronic devices with unique anisotropic nature.

## 4. Experimental Section

TiS$_3$ Preparation: The TiS$_3$ microcrystals were synthesized by sulfuration of titanium powder which is vacuum sealed in an ampule with sulfur powder (>75% atomic sulfur) and heated to 500ºC. After 20 hours of growth, the ampule is cooled down to ambient conditions.[10]

Characterizations: The atomic force microscopy (AFM) of the devices was performed with a commercial head and software from Nanotec. The as-grown TiS$_3$/Si heterostructure devices were annealed at 200 ºC for 2h in vacuum to improve the contact for the devices. The optical properties of the nanosheets have been studied with a home-built micro-reflectance/transmittance spectroscopy setup, described in detail in Ref.[12]. The transmittance (Figure 1) and absorbance (Figure 3) data have been obtained through differential measurements. Two spectra are acquired: one on the bare substrate ($I_0$) and another one on the TiS$_3$ sample ($I$). The transmittance ($T$) is then calculated as $T = I/I_0$ and the absorbance ($A$) is $A = -\log_{10} T$. Therefore, the absorption of the PDMS substrate is accounted for in the calculation of these magnitudes. These devices were characterized by a semiconductor analyzer Keithley 2450 and a home-built probe station.

**Acknowledgements**

We acknowledge funding from the EU Graphene Flagship (Grant Graphene Core2 785219) and from the European Research Council (ERC) under the European Union's Horizon 2020 research and innovation programme (grant agreement n° 755655, ERC-StG 2017 project 2D-TOPSENSE). RF acknowledges support from the Netherlands Organisation for Scientific Research (NWO) through the research program Rubicon with project number 680-50-1515. DPdL acknowledges support from the MINECO (program FIS2015-67367-C2-1-P). YN acknowledges the grant from the China Scholarship Council (File NO. 201506120102). WCJ acknowledges the Program for the Top Young and Middle-aged Innovative Talents of Harbin Institute of Technology. MIRE group acknowledges financial support from MINECO-FEDER (MAT2015-65203-R).

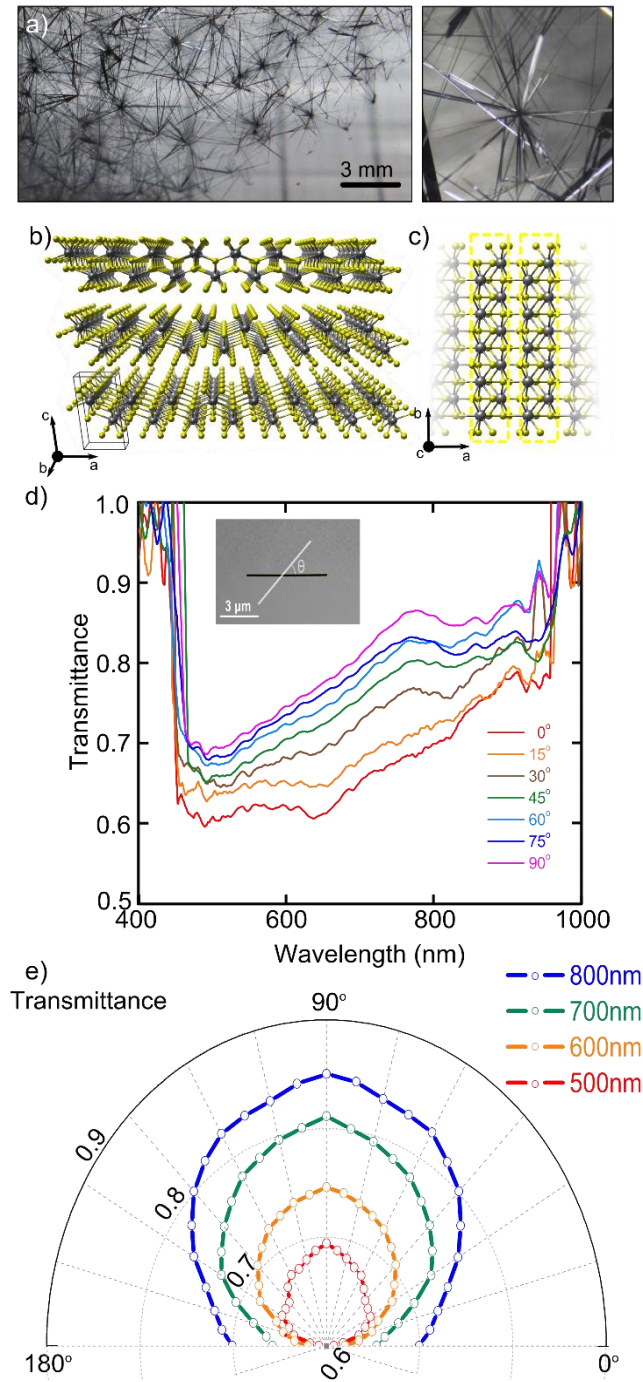

**Figure 1. a**) Photograph of TiS$_3$ and magnified photograph of TiS$_3$ ribbons. **b**) Artistic representation of the crystal structure of TiS$_3$. The gray spheres represent the Ti atoms and the yellow spheres represent the S atoms. The unit cell is indicated by solid black lines. **c**) Artistic representation of the TiS$_3$ unit cell. **d**) Differential transmittance spectra of a TiS$_3$ nanoribbon deposited onto PDMS measured as a function of the polarization angle of the transmitted radiation. Inset: Photograph of a TiS$_3$ ribbon (the white solid line defines the polarization angle). **e**) Polar plot of the transmittance of a TiS$_3$ nanoribbon at different wavelengths.

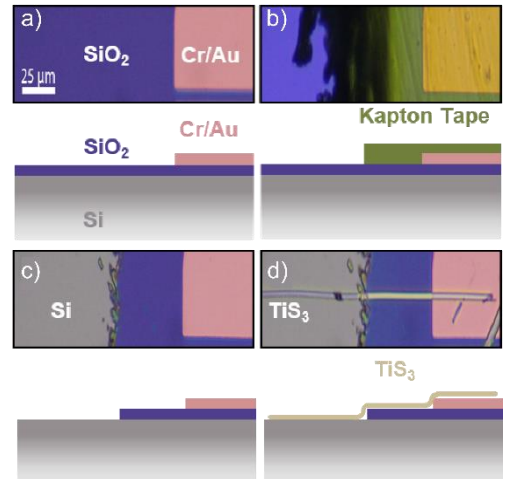

**Figure 2.** Schematic diagram of preparation process of $TiS_3$/Si p-n junction. **a)** Pre-Patterned Cr/Au electrode on $SiO_2$/Si Substrate. **b)** Covering electrode with kapton tape. **c)** Etching the $SiO_2$ from Si Substrate. **d)** Preparing the p-n junction by transferring a $TiS_3$ nanoribbon onto the partially exposed Si substrate.

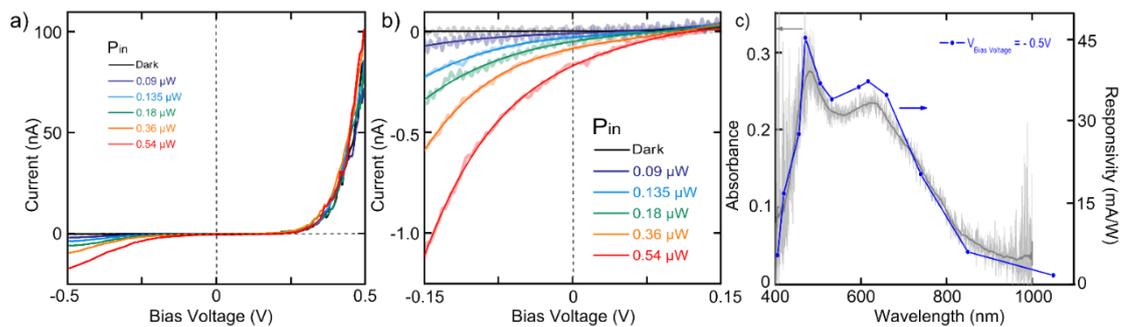

**Figure 3. a)** Current vs. bias voltage characteristics of the device in dark and upon illumination with increasingly optical power. **b)** zoomed in plot of current vs. bias voltage characteristics of the device. **c)** Responsivities (blue line) for different illumination wavelengths with an incident optical power of 0.54 μW and a bias voltage of -0.5 V. The responsivities are compared with the micro-absorbance spectrum (grey line) of $TiS_3$ on PDMS.

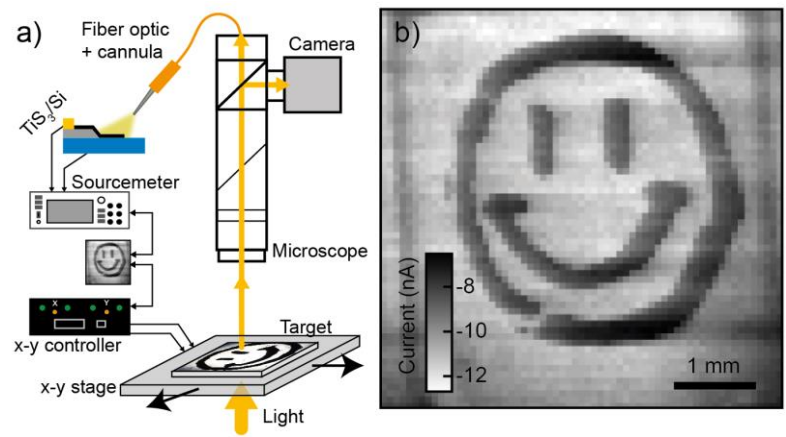

**Figure 4. a)** Schematic of the experimental system for single pixel camera. **b)** The recovered image by the single-pixel camera of the TiS$_3$/Si photodetector with 64 · 64 pixels at -1V.

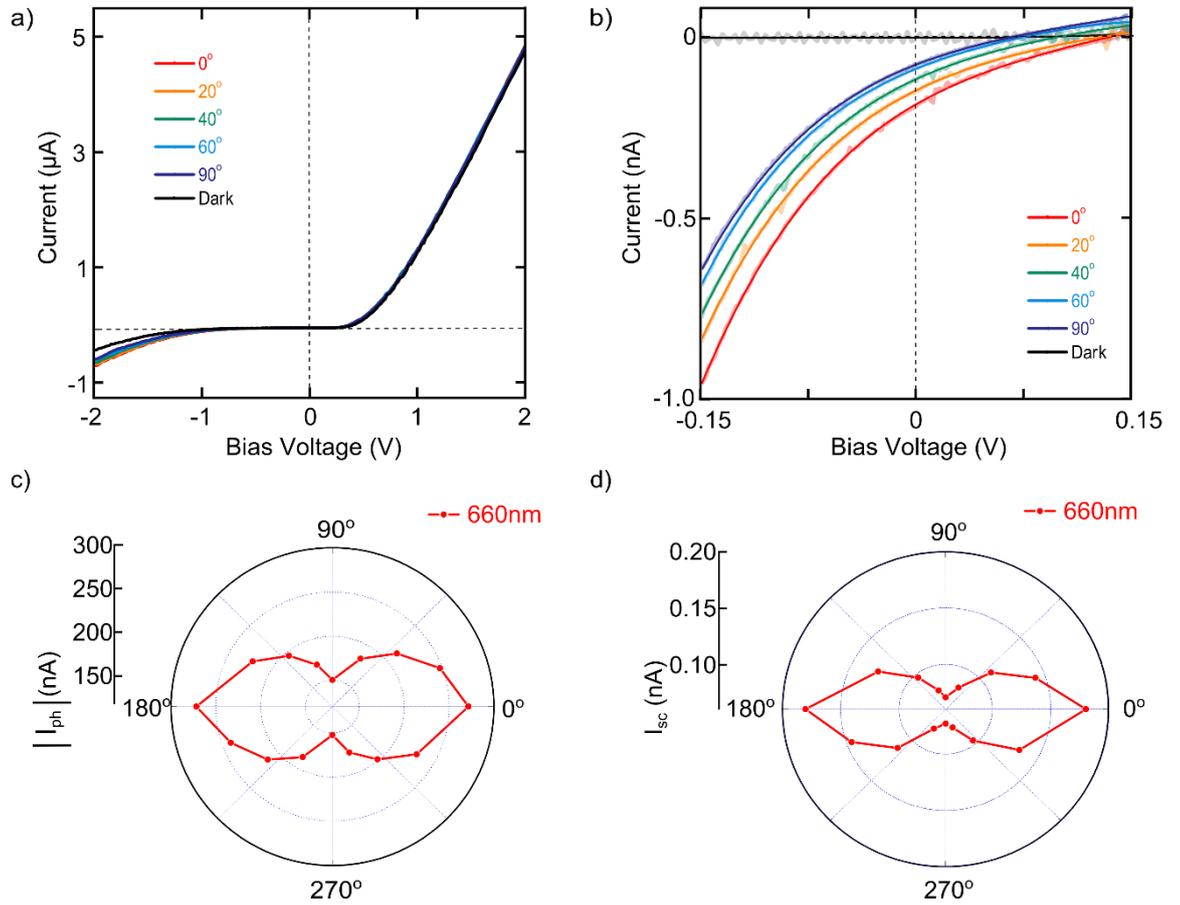

**Figure 5. a)** Current vs. bias voltage characteristics of the TiS$_3$/Si device upon illumination with increasing polarization angle. **b)** A zoomed in plot around zero bias to show the photovoltaic effect upon illumination. **c)** Polar plot of the absolute value of photocurrent at a bias voltage of -2 Volts. **d)** Polar plot of the absolute value of the short circuit current.

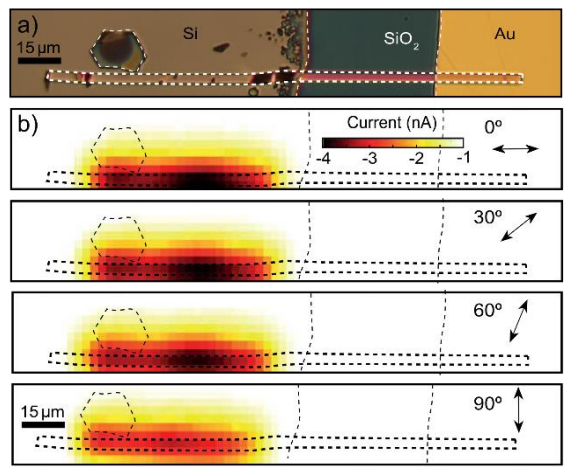

**Figure 6. a)** Optical image of the p-n heterojunction used for photocurrent mapping. **b)** Scanning photocurrent map of the p-n heterojunction at different polarization angles. The dashed lines outline the important spatial features of the device.

**Table 1.** Performance Comparison of our TiS3/Si devices with reported anisotropic photodetectors

| Device | Measure ment | Self-powered | Cut-Off Wavelength nm | $R^{(a)}$ mA/W | $D^{*}$ [(b)] cm Hz$^{1/2}$W$^{-1}$ | Respons e time ms | Anisotropy ratio | Ref |
|---|---|---|---|---|---|---|---|---|
| TiS$_3$/Si heterojunction | 660nm V= - 0.5 V/ 0 V | Yes | >1050 | 34.8 / 0.33 | $1.04 \cdot 10^{10}$ / $2.5 \cdot 10^8$ | <20 | 0.29 / 0.44 | This work Main text |
| | | | | 18.3 / 0.31 | $1.3 \cdot 10^{10}$ / $2.9 \cdot 10^8$ | <20 | 0.21 / 0.49 | Sample 2 |
| | | | | 8.5 / 2.63 | $3.36 \cdot 10^9$ / $1.9 \cdot 10^9$ | <20 | 0.26 / 0.38 | Sample 3 |
| | | | | 18.9 / 3 | $6.63 \cdot 10^9$ / $2.65 \cdot 10^9$ | <20 | 0.24 / 0.64 | Sample 4 |
| Black phosphorus | 1200 nm V= 0.1 V | No | 3750 | 0.35 | $4.13 \cdot 10^7$ [(*)] | 0.04 | 0.56 | [8b] |
| Black phosphorus | 785 nm V= -10 V | No | N/A | N/A | N/A | N/A | 0.3 | [8d] |
| ReS$_2$ | 405 nm V= 1 V | No | 800 | $1.2 \cdot 10^6$ | $5 \cdot 10^{11}$ [(*)] | 98 | 0.5[(*)] | [8c] |
| ReSe$_2$ | 633 nm V$_{ds}$= 0.5 V | No | 1000 | 1.5 | $1.19 \cdot 10^8$ [(*)] | 2 | 0.33[(*)] | [8c] |
| GeS | 500 nm V$_{ds}$= 4 V 373K | No | 800 | $2.4 \cdot 10^6$ | $1.5 \cdot 10^{14}$ | N/A | >0.9[(*)] | [17a] |
| InP | N/A | No | N/A | $3 \cdot 10^6$ | N/A | N/A | 0.96 | [19a] |
| SWNT | 970 nm V$_{ds}$= 1 V V$_g$= -1.5 V | No | 980 | N/A | N/A | N/A | 0.22[(*)] | [19b] |
| Black phosphorus/MoS$_2$ | 532 nm V= 2 V | Yes | >1550 | 170 | $4.9 \cdot 10^8$ [(*)] | N/A | 0.43[(*)] | [8a] |
| Black phosphorus/WSe$_2$ | 1550 nm V= 0.5 V | No | >1550 | $5 \cdot 10^2$ | $10^{10}$ | 0.8 | 0.71[(*)] | [17b] |

Table Footnote, N/A: not applicable, R(a): Responsivity of the device and D* (b): Detectivity of the device with the conditions specified in the column 'Measurements', (*): The value calculated from the data extracted from the reference but not stated directly in the cited article. Note that in some columns we display two values for our devices, they correspond to the values obtained under V= - 0.5 V and V= 0 V, respectively.

# Supporting Information

**Polarization-sensitive and broadband photodetection based on a mixed-dimensionality TiS₃/Si p-n junction**

Yue Niu[1,2], Riccardo Frisenda[2]*, Eduardo Flores[3], Jose R. Ares[3], Weicheng Jiao[1], David Perez de Lara[2], Carlos Sanchez[3,4], Rongguo Wang[1], Isabel J. Ferrer[3,4]* and Andres Castellanos-Gomez[5]*

[1]Science and Technology on Advanced Composites in Special Environments Laboratory, Harbin Institute of Technology, 150080, Harbin, P. R. China

[2]Madrid Institute for Advanced Studies (IMDEA Nanociencia). Ciudad Universitaria de Cantoblanco Calle Faraday 9, 28049 Madrid, Spain
E-mail: riccardo.frisenda@imdea.org

[3]Materials of Interest in Renewable Energies Group (MIRE Group). Dpto. de Física de Materiales, Universidad Autónoma de Madrid, UAM, Campus de Cantoblanco, E-28049 Madrid, Spain.
E-mail: isabel.j.ferrer@uam.es

[4]Instituto Nicolás Cabrera, Universidad Autónoma de Madrid, UAM, Campus de Cantoblanco E-28049 Madrid, Spain.

[5]Materials Science Factory, Instituto de Ciencia de Materiales de Madrid (ICMM-CSIC), Madrid, E-28049, Spain.
E-mail: andres.castellanos@csic.es

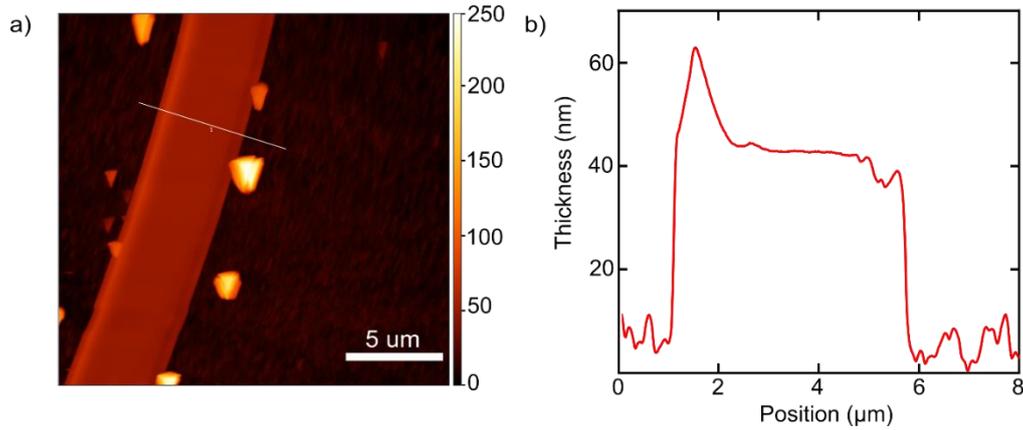

**Supplementary Figure 1. a)** Atomic force microscopy image of part of the junction region. **b)** Atomic force microscopy line profile that indicates the thickness.

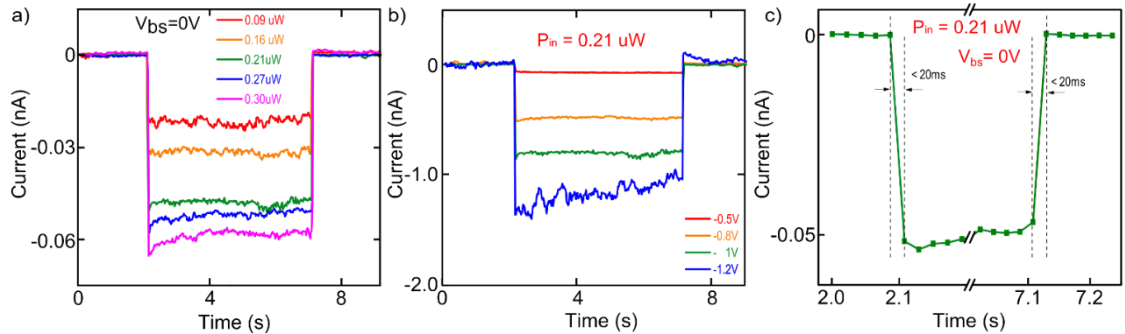

**Supplementary Figure 2. a)** Current response at 0V bias voltage under a 50 Hz mechanically modulated optical excitation at 660 nm with increasingly power. **b)** Current response with increasing bias voltage under a 50 Hz mechanically modulated optical excitation at 660 nm with decreasingly bias voltage. **c)** Zoom in a single switching cycle at 50 Hz frequency.

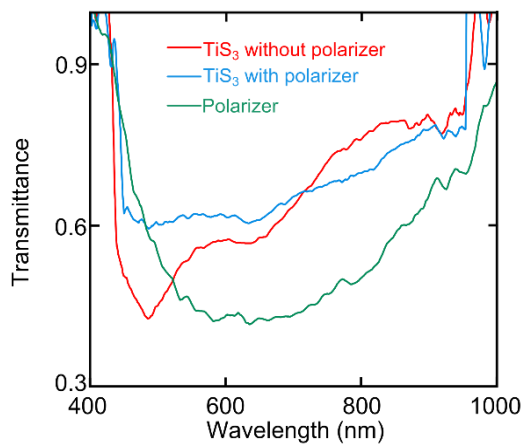

**Supplementary Figure 3.** Transmittance of TiS$_3$ without polarizer (red), TiS$_3$ with polarizer (blue) and polarizer (green).

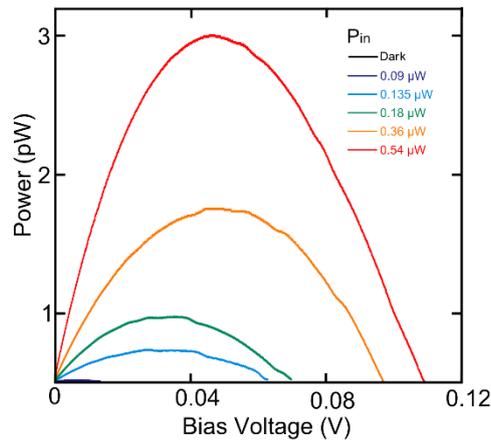

**Supplementary Figure 4.** Photovoltaic generated power, calculated from the current vs. bias voltage characteristics.

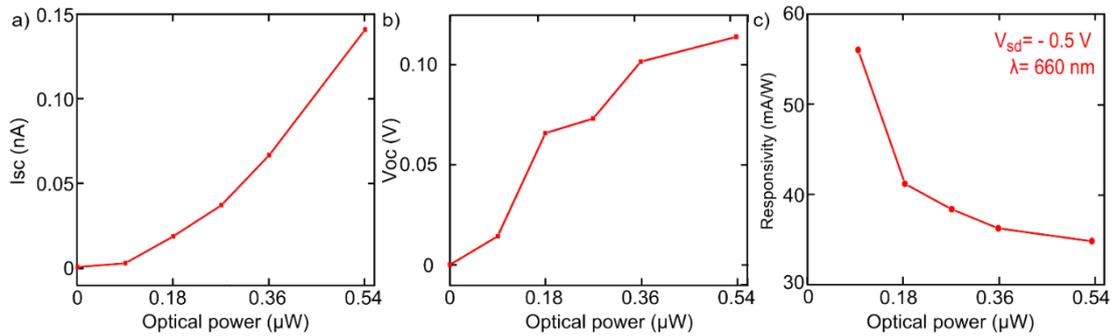

**Supplementary Figure 5. a)** Power dependence of the short circuit current. **b)** Power dependence of the open circuit voltage. **c,** Power dependence of the responsivity.

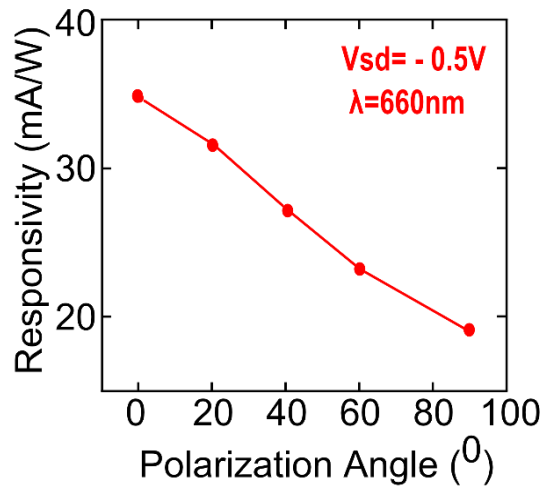

**Supplementary Figure 6** Polarization dependence of responsivity at a bias voltage of -0.5 Volts.

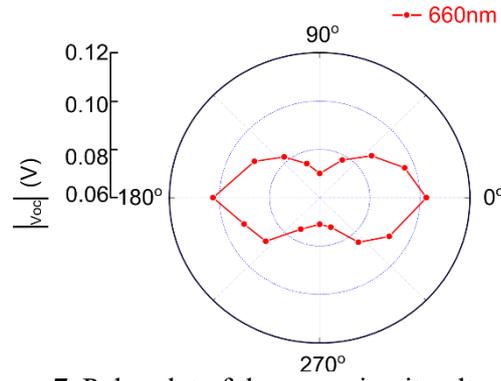

**Supplementary Figure 7.** Polar plot of the open circuit voltage ($\lambda$ = 660 nm).

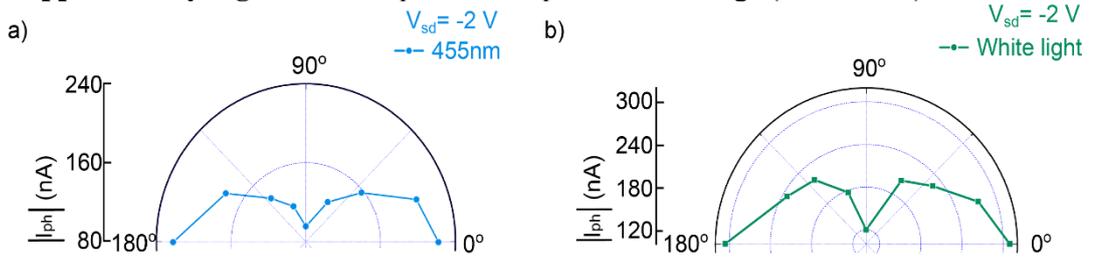

**Supplementary Figure 8. a)** Polar plot of the absolute value of photocurrent ($\lambda$ = 455 nm). **b)** Polar plot of the absolute value of photocurrent (White light).

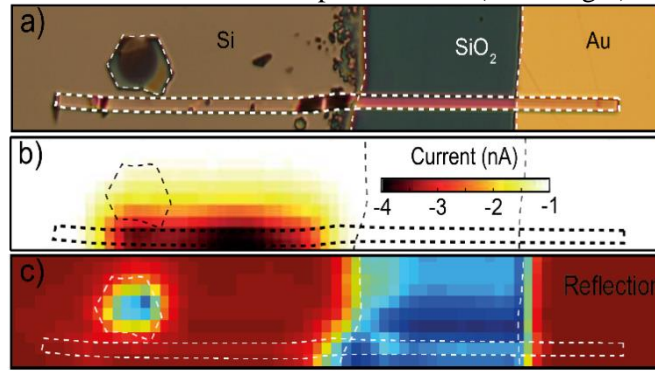

**Supplementary Figure 9.** Optical image of the p-n heterojunction used for photocurrent mapping and reflection mapping.

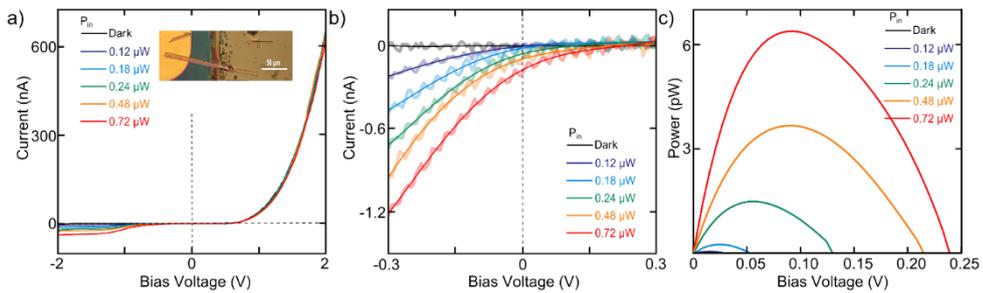

**Supplementary Figure 10. a)** Current vs. bias voltage characteristics of the 2nd device in dark and upon illumination with increasingly optical power. Inset: Optical image of the p-n heterojunction. **b)** A zoomed in plot of current vs. bias voltage characteristics of the device in dark and upon illumination with increasingly optical power. **c)** Photovoltaic generated power, calculated from the current vs. bias voltage characteristics

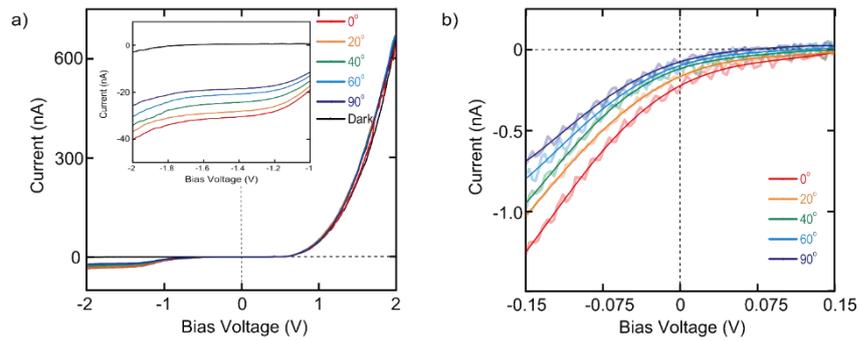

**Supplementary Figure 11. a)** Current vs. bias voltage characteristics of the 2nd device upon illumination with increasing polarization angle. Inset: A zoomed in plot around a bias voltage of -2 Volts. **b)** A zoomed in plot around zero bias to show the photovoltaic effect upon illumination.

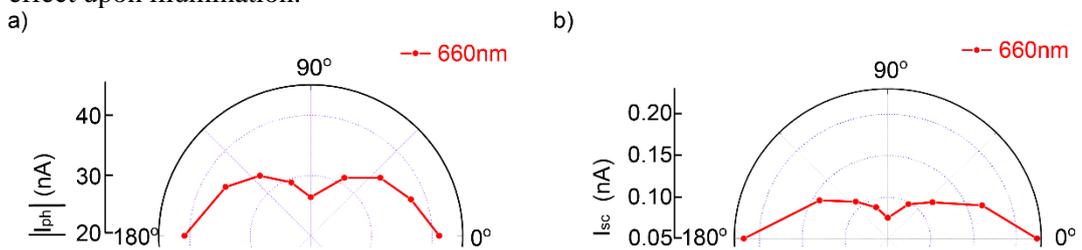

**Supplementary Figure 12. a)** Polar plot of the absolute value of photocurrent of the 2nd device at a bias voltage of -2 Volts. **b)** Polar plot of the absolute value of the 2nd device of the short circuit current.

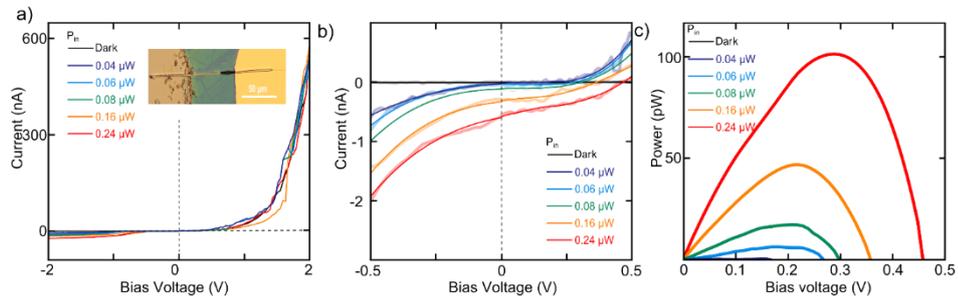

**Supplementary Figure 13. a)** Current vs. bias voltage characteristics of the 3rd device in dark and upon illumination with increasingly optical power. Inset: Optical image of the p-n heterojunction. **b)** A zoomed in plot of current vs. bias voltage characteristics of the device in dark and upon illumination with increasingly optical power. **c)** Photovoltaic generated power, calculated from the current vs. bias voltage characteristics

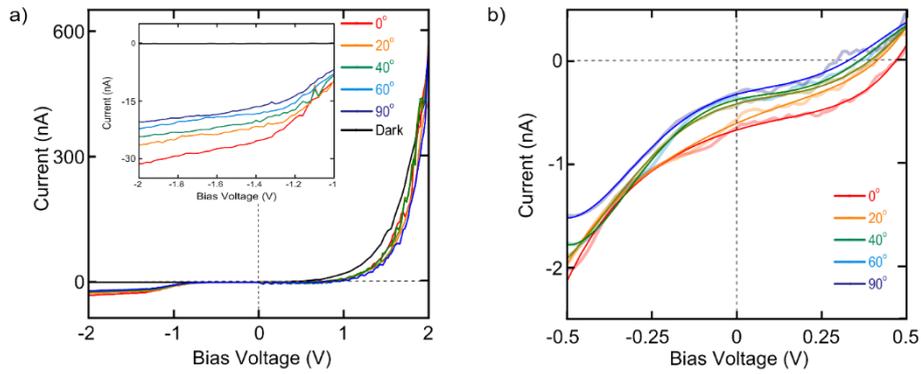

**Supplementary Figure 14. a)** Current vs. bias voltage characteristics of the 3rd device upon illumination with increasing polarization angle. Inset: A zoomed in plot around a bias voltage of -2 Volts. **b)** A zoomed in plot around zero bias to show the photovoltaic effect upon illumination.

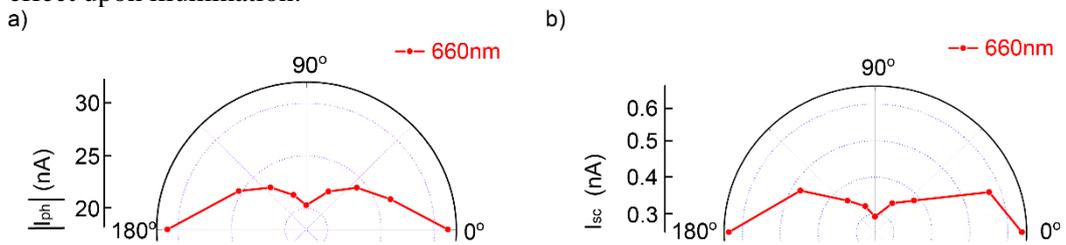

**Supplementary Figure 15. a)** Polar plot of the absolute value of photocurrent of the 3rd device at a bias voltage of -2 Volts. **b)** Polar plot of the absolute value of the 3rd device of the short circuit current.

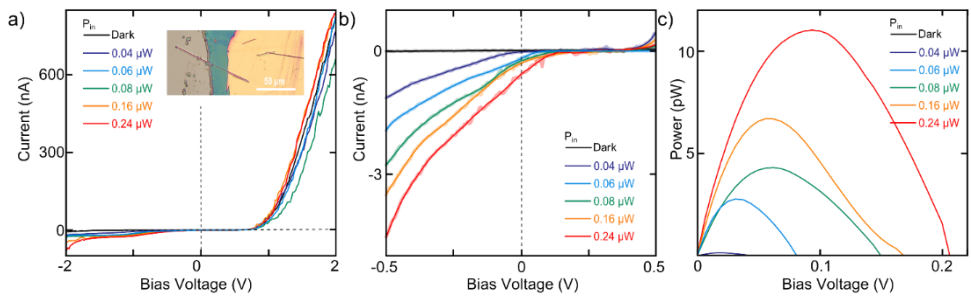

**Supplementary Figure 16. a)** Current vs. bias voltage characteristics of the 4th device in dark and upon illumination with increasingly optical power. Inset: Optical image of the p-n heterojunction. **b)** A zoomed in plot of current vs. bias voltage characteristics of the device in dark and upon illumination with increasingly optical power. **c)** Photovoltaic generated power, calculated from the current vs. bias voltage characteristics

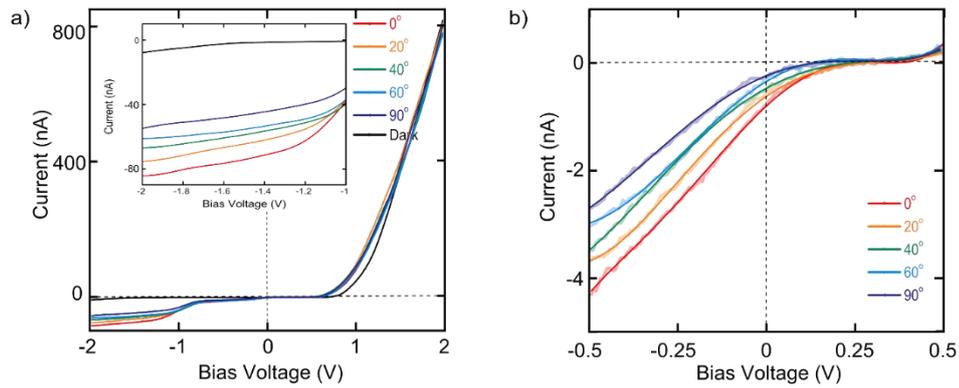

**Supplementary Figure 17. a**) Current vs. bias voltage characteristics of the 4th device upon illumination with increasing polarization angle. Inset: A zoomed in plot around a bias voltage of -2 Volts. **b**) A zoomed in plot around zero bias to show the photovoltaic effect upon illumination.

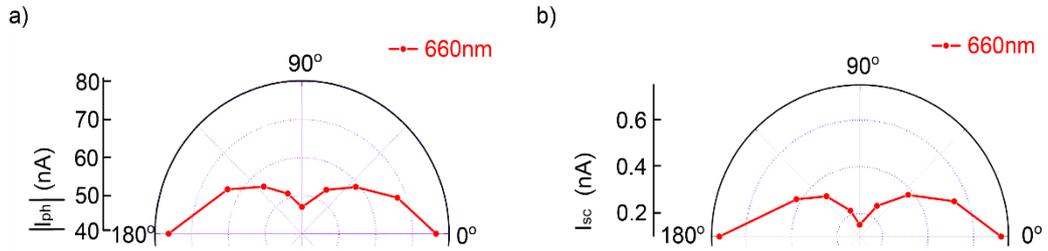

**Supplementary Figure 18. a**) Polar plot of the absolute value of photocurrent of the 4th device at a bias voltage of -2 Volts. **b**) Polar plot of the absolute value of the 4th device of the short circuit current.